\documentclass[prl,superscriptaddress,showpacs,amsmath,amsfonts,twocolumn,floatfix]{revtex4}
\usepackage[final]{graphicx}
\usepackage[sort&compress]{natbib}

\newlength\figurewidth
\makeatletter
\@ifxundefined\draft@sw{\@booleanfalse\draft@sw}{}
\draft@sw{%
\AtBeginDocument{\setlength\figurewidth{.7\linewidth}}
}{%
\AtBeginDocument{\setlength\figurewidth{.9\linewidth}}
}
\makeatother

\def\kB{k_{\text{B}}}

\def\l{\text{l}}
\def\s{\text{s}}

\begin{document}

\title{A Double-Transition Scenario for Anomalous Diffusion in Glass-Forming Mixtures}
\date\today
\def\mpdlr{\affiliation{%
  Institut f\"ur Materialphysik im Weltraum,
  Deutsches Zentrum f\"ur Luft- und Raumfahrt (DLR),
  51170 K\"oln, Germany}}
\def\unikn{\affiliation{%
  Fachbereich Physik, Universit\"at Konstanz,
  78457 Konstanz, Germany}}
\author{Th.~Voigtmann}\mpdlr\unikn
\author{J.~Horbach}\mpdlr

\begin{abstract}
We study by molecular dynamics computer simulation a binary
soft-sphere mixture that shows a pronounced decoupling
of the species' long-time dynamics.
%in the time scales connected to self diffusion of its species,
Anomalous, power-law-like diffusion of small particles arises,
that can be understood as a precursor of a double-transition
scenario, combining a glass transition and a separate small-particle
localization transition. %, as predicted by mode-coupling theory.
%The small-particle dynamics is the result of an intricate interplay
%between free-volume caging and percolation dynamics.
Switching off small-particle excluded-volume constraints slows down,
rather than enhances, small-particle transport.
The data are contrasted with results from the mode-coupling theory
of the glass transition.
\end{abstract}

\maketitle

%
%\section{Introduction}
%

Transport properties in disordered media are important in a wide
range of applications from biophysics to geosciences.
Intriguing behavior arises from `fast' species moving through a dense
host system, such as power-law-like dynamical conductivities of
ion-conducting melts \cite{Jonscher.1977}.
Likewise, `anomalous diffusion' appears in many amorphous media:
mean-squared displacements (MSD) that grow like
$\delta r^2\sim t^\mu$ (with some positive $\mu<1$) over large time windows,
instead of obeying Einstein's law for ordinary diffusion ($\mu=1$),
are seen in biophysical
tracer experiments investigating cellular environments
\cite{Feder.1996,Wong.2004,Banks.2005},
in zeolites \cite{Hahn.1996,Sholl.1997}, gels \cite{Netz.1995,Babu.2008},
amorphous semiconductors and photoconductors \cite{Scher.1975}, or
specially confined
colloidal suspensions \cite{Wei.2000,Lutz.2004,Lin.2005}.

These systems can be thought of as mixtures composed of a small (fast)
species and slow (big) host particles providing
%an almost frozen background that forms
a highly complex confining structure (called
`molecular crowding' in biophysical literature).
%Taking aside anomalous diffusion of polymer monomers, % \cite{Shusterman.2004},
One way to deal with this, is to
simplify the discussion to stochastic lattice gases and single tracers moving
in a random environment \cite{Maass.1991,Saxton.1994,Metzler.2000}, invoking as a
reference point the single-file diffusion of non-overtaking particles,
$\delta r^2\sim t^{1/2}$ \cite{Harris.1965,Levitt.1973,Lizana.2008}.
Such modeling obviously leaves out two aspects:
the dynamics leading to a time-scale separation in the first place, and
interactions among the carrier particles.

In order to highlight the remarkable features arising from dynamical
many-body effects in anomalous diffusion, we investigate a binary,
disparate-size soft-sphere mixture.
%Ignoring all non-entropic interactions,
We show how anomalous diffusion can be interpreted as a
high-density phenomenon, specifically as the approach to a double-glass
transition.
Many-body interactions manifest themselves in a striking way in the
dynamics of the small species: \emph{releasing} excluded-volume constraints,
their mobility is \emph{reduced} at long times, rather than enhanced.

The appearance of two kinds of glasses -- one where both particle species
freeze, one where the smaller one stays mobile -- has been predicted
\cite{Sjoegren1986,Bosse1987,Bosse1995} using mode-coupling theory
of the glass transition (MCT) \cite{Goetze1991b}, and indicated in
colloidal experiments \cite{Imhof1995a,Imhof1995b} and molecular-dynamics (MD)
simulations \cite{Moreno2006b}. MCT qualitatively explains fast-ion diffusion
in sodium silicate melts \cite{Voigtmann2006} as a precursor of this scenario.
The two transitions have different microscopic origins: while the slow dynamics
of the larger species is dominated by caging on the nearest-neighbor scale,
the single-particle dynamics of the smaller species exhibits a
continuously diverging localization length. This latter leads to the
appearance of power-law-like anomalous diffusion.
A similar transition also appears in a recent extension of MCT
where big particles are immobile from the outset
\cite{Krako2005,Krako2006}.

The exemplary off-lattice model for particle localization is the
Lorentz gas (LG), a single classical point particle
moving between randomly distributed, fixed hard-sphere obstacles. At
a critical obstacle density, the particle undergoes a localization transition
understood as a critical dynamic phenomenon \cite{Havlin2002}.
Close to the transition, a power-law asymptote for the
mean-squared displacement is explained by continuum percolation
theory, as demonstrated in recent extensive simulations
\cite{Hoefling.2006,Hoefling.2007,Hoefling.2008}.
We shall embark on the subtle connection between the LG
and true binary mixtures below.
%The connection between the LG and the MCT scenarios is richer than
%anticipated due to collective effects, as we shall point out below.

%\section{Model System}
\begin{figure}
\includegraphics[width=0.95\figurewidth]{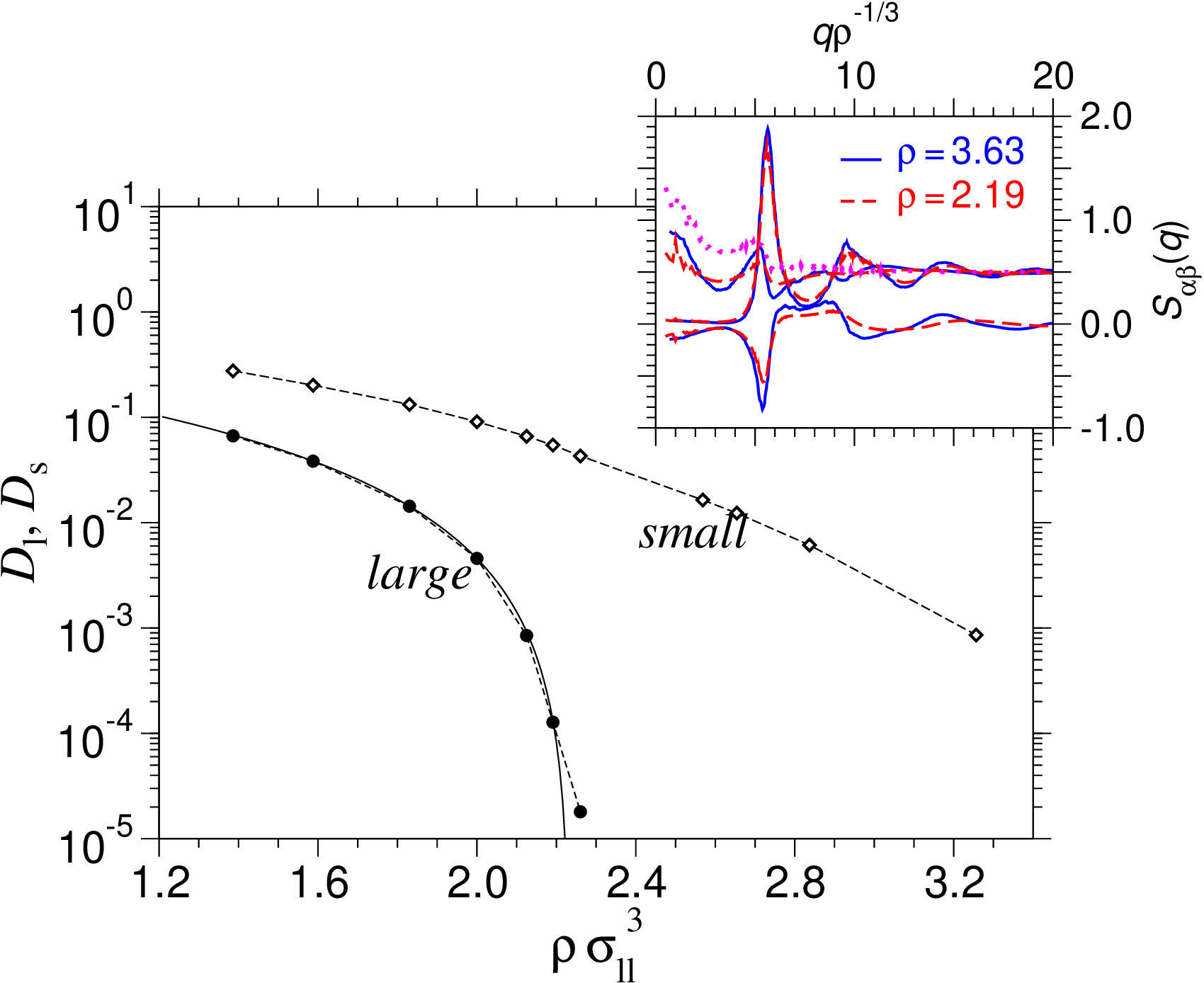}
\caption{\label{fig:diff}
  MD-simulated self-diffusion coefficients for small ($D_\s$)
  and large particles ($D_\l$) in a disparate-sized binary mixture.
  The solid line is a power-law fit
  $\propto (\rho_c - \rho)^{\gamma}$, where $\rho_c=2.23$ and
  $\gamma=2.1$. % Dashed lines are guides to the eye.
  Inset: partial static structure factors $S_{\alpha\beta}(q)$
  at two different densities as indicated, as functions of $q\rho^{-1/3}$.
  The dotted line is $S_{\s\s}(q)$ for non-interacting small particles
  at $\rho=2.19$.
}
\end{figure}

We performed molecular-dynamics (MD) simulations of an equimolar binary
mixture of purely repulsive soft spheres, with interaction potential
$V_{\alpha\beta}(r)=4\epsilon_{\alpha\beta}[(r/\sigma_{\alpha\beta})^{-12}
-(r/\sigma_{\alpha\beta})^{-6}]+\epsilon_{\alpha\beta}$
for $r<r_-=2^{1/6}\sigma_{\alpha\beta}$ (zero else), $\alpha,\beta\in(\l,\s)$.
Diameters are chosen additively, $\sigma_{\alpha\beta}=(\sigma_{\alpha\alpha}
+\sigma_{\beta\beta})/2$, $\sigma_{\l\l}$ setting the unit of
length, and $\sigma_{\s\s}=0.35$.
Nonadditive energetic interactions further decouple the species,
$\epsilon_{\l\l}=\epsilon_{\s\s}=1$ but
$\epsilon_{\s\l}=0.1$.
%This is roughly equivalent to introducing a nonadditive diameter with
%$\sigma_{\s\l}=\epsilon_{\s\l}^{1/12}(\sigma_{\l\l}+\sigma_{\s\s})/2
%\approx0.29$.
We set the temperature $\kB T=2/3$, and all masses equal,
$m_\l=m_\s=1$: no time-scale separation exists between
the two species at short times.

The smoothened potential
$V(r)\times[(r-r_-)/h]^4/[1+(r-r_-)/h]^4$
with $h=0.005\sigma_{\l\l}$ provides continuity of energy and forces
at the cutoff $r_-$.  Newton's equations of motion were integrated
for $N_\l=N_\s=1000$ particles with the velocity form
of the Verlet algorithm (time step $\delta t = 0.005/\sqrt{48}$
in units $t_0=[m_\l\sigma_{\l\l}^2/\epsilon_{\l\l}]^{1/2}$).  To avoid
crystallization, big-particle diameters were sampled equidistantly
from the interval $\sigma_{\l\l}\in[0.85,1.15]$,
retaining $\sigma_{\s\l}=(1+\sigma_{\s\s})/2$.
%so that the small particles
%essentially `see' a single size of big ones.
At each number density $\rho$, four independent runs were performed.
Up to $\rho\le2.296\,\sigma_{\l\l}^{-3}$, the system was fully equilibrated,
requiring equilibration runs over at least $10^6$ and up to
$2\times10^8$ time steps, followed by production runs of the same length.
During equilibration, temperature was held constant by coupling the
system periodically to a stochastic heat bath; production runs were
done in the microcanonical ensemble.
At the highest density $\rho=4.215\, \sigma_{\l\l}^{-3}$, over $10^9$
time steps were performed.  No runs showed signs of demixing or equilibrium
phase transitions.

%\section{MD Results}

% FIG: self-diffusion constants
%
%

Figure~\ref{fig:diff} displays the self-diffusion constants $D_{\alpha}$
obtained from the simulated mean-squared displacement (MSD),
$\delta r_\alpha^2(t)=\langle(\vec r^s_\alpha(t)-\vec r^s_\alpha(0))^2
\rangle$ for a singled-out particle at $\vec r^s_\alpha(t)$
via the Einstein relation,
$\delta r_\alpha^2(t\to\infty)\sim6D_\alpha t$, where possible.
The diffusion coefficients show a
decoupling between the motion of large and small particles which becomes
more pronounced with increasing density, due to a faster slowing down in
$D_\l$ than in $D_\s$. At $\rho=2.296$,
%(the highest density where the system could be fully equilibrated),
$D_\s$ is about 2.5 orders of magnitude higher than $D_\l$,
and at $\rho\ge2.568$, big-particle diffusion has ceased over the entire
simulation time window. Yet, the small-particle MSD still retains
a diffusive regime, allowing us to extract $D_\s>0$ up to
$\rho=3.257$. Also shown in
Fig.~\ref{fig:diff} is a fit to $D_\l$ by a power law as predicted
by MCT, $D\sim D_0(\rho_c - \rho)^{\gamma}$. The fit yields $\rho_c=2.23$
for the critical density and $\gamma=2.1$ for the critical exponent.
A similar fit to $D_\s$ is not satisfying and yields much larger
$\gamma$ and $\rho_c$. $D_\s(\rho)$ also clearly differs from an
Arrhenius law, although a fit with two exponentials would be possible.

% FIG: S(q)
%
%\begin{figure}
%\includegraphics[width=0.9\figurewidth]{figures/sofq2}
%\caption{\label{fig:sq}
%  Partial static structure factors $S_{\alpha\beta}(q)$ for the binary
%  soft-sphere mixture at two different densities as indicated, as functions
%  of $q\rho^{-1/3}$.
%}
%\end{figure}

The slowing down visible in Fig.~\ref{fig:diff} and discussed in the following
is purely dynamic; no essential changes in the static structure of the system
were observed, despite the drastic compression employed. This is demonstrated
by the partial static structure factors
$S_{\alpha\beta}(q)=\langle\sum_{jk}\exp[-i\vec q\cdot(\vec r_{\alpha,j}
-\vec r_{\beta,k})]\rangle$, showing little change with density if plotted
as functions of $q^*=q\rho^{-1/3}$ to eliminate a trivial change in length
scale (inset of Fig.~\ref{fig:diff}).

% FIG: MSD simulation large + small
%
\begin{figure}
\includegraphics[width=0.9\figurewidth]{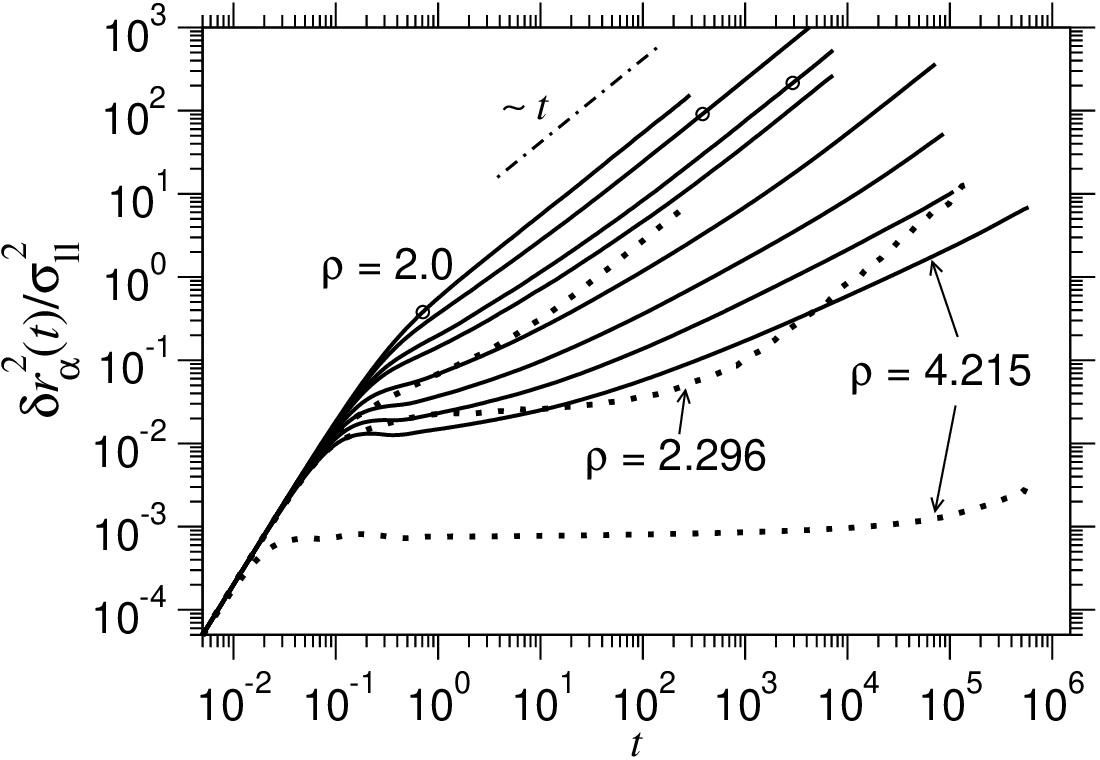}
\caption{\label{fig:msdsim}
  Mean-squared displacements (MSD), $\delta r_\alpha^2(t)$ for the
  large (dotted) and small (solid lines) particles in the
  simulated binary mixture.
  Densities shown are $\rho\sigma_{\l\l}^3=2.0$, $2.296$, and $4.215$ for
  $\alpha=\l$ and
  $\rho\sigma_{\l\l}^3= 2.0$, $2.296$, $2.654$, $2.837$, $3.257$, $3.627$,
  $3.906$, and $4.215$ for $\alpha=\s$. Symbols mark the crossover to
  ordinary diffusion, where $\delta r_\s^2(t)\sim t$ (slope indicated
  by the dash-dotted line).
}
\end{figure}

The MSD corresponding to these transport coefficients are shown in
Fig.~\ref{fig:msdsim}. For the big particles, we observe
a standard glass-transition scenario: a two-step process gives rise to
a plateau over an increasingly large time window,
crossing over to diffusion at increasingly large time, and at a length
scale associated with dynamic nearest-neighbor cageing, typically at about
$10\%$ of a particle diameter (Lindemann's criterion).
Indeed, from the plateau of the $\rho=2.296$ curve one reads off the
corresponding cage
localization length $r^c_\l=\sqrt{\delta r^2_\l/6}\approx0.06\,\sigma_{\l\l}$,
which decreases at larger $\rho$ due to compression.
Although this could technically be called anomalous diffusion, we reserve
that term for the behavior shown by the small particles:
Around $\rho_c$, the small-particle MSD behave quite
differently, with no sign of a two-step glassy dynamics.
Instead, they show subdiffusive
growth and cross over to ordinary diffusion at increasingly large length
and time scales when increasing $\rho$. This indicates that nearest-neighbour
cageing is not the dominant mechanism
for their slowing down. The sub\-diffusive regime can be
described by power-law variation, $\delta r_\s^2(t)\propto t^\mu$ with
some effective $0<\mu<1$ that seems to decrease with increasing density.

The small-particle dynamics qualitatively agrees with previous MD results
\cite{Moreno2006b}. It also agrees with the dynamics found in the Lorentz gas
\cite{Hoefling.2006,Hoefling.2007,Hoefling.2008}.
There, subdiffusive growth with apparent
density-dependent exponents $\mu$ is due to the approach to an
asymptotic power law, $\delta r_\s^2(t)\sim t^x$, that extends to
$t\to\infty$ at the localization critical point. Careful simulations
\cite{Hoefling.2006} could establish $x=2/6.25$ for the LG. To
estimate a critical exponent $x$ from
Fig.~\ref{fig:msdsim} is tempting, but preasymptotic corrections
render it impossible.
It appears that a description of our data in terms of the LG asymptote
is not convincing.

% FIG: MSD low and high density, comparison WCA and LG-WCA
%
\begin{figure}
\includegraphics[width=0.9\figurewidth]{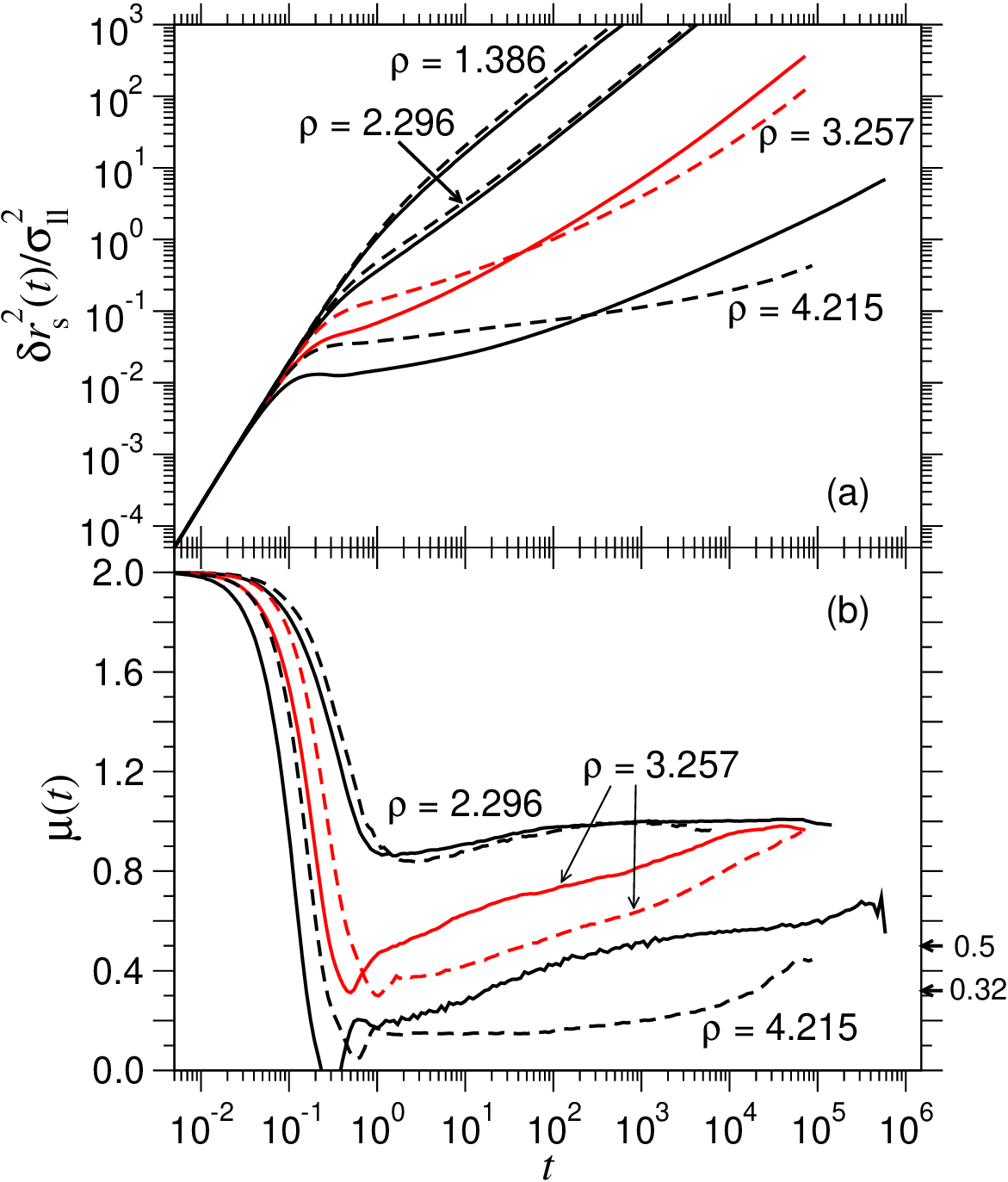}
\caption{\label{fig:msdlgsim}
  (a) Small-particle MSD with (solid lines) and
  without (dashed) interactions between small particles,
  densities as indicated.
  (b) Logarithmic derivative
  $\mu(t)=d[\log\delta r_\s^2(t)]/d(\log t)$.
}
\end{figure}

Our binary mixture differs from the LG \textit{inter alia} through the
finite density of interacting small particles. To establish the effect of
this distinction, we switch off interactions among small particles,
setting $\epsilon_{\s\s}=0$ while keeping their number constant.
Within simulation accuracy, structure and
dynamics of the big particles are unchanged
in this `transparent-small' mixture.

Figure~\ref{fig:msdlgsim} compares the small-particle MSD
of the two systems. Initially, the transparent small particles show weaker
localization, intuitively expected as they have larger free volume available.
This trend prevails at low densities.
Surprisingly, at high $\rho$, switching \emph{off}
interactions
%between the small particles
leads to significantly \emph{slower} diffusion
compared with the fully interacting case as $t\to\infty$.

This is emphasized by the lower panel of Fig.~\ref{fig:msdlgsim}:
the effective exponent $\mu(t)=d[\log\delta r_\s^2(t)]/d(\log t)$
crosses over from $\mu(0)=2$ (ballistic short-time motion) to
$\mu(\infty)=1$ for ordinary diffusion or $\mu(\infty)=0$ for
arrested particles. For the LG model, $\mu(t)\approx x$ for increasingly
large time windows close to the localization treshold. No clear plateaus
are seen in our data, but switching off small--small interactions
at fixed density $\rho$
systematically reduces $\mu(t)$ at long times.
%suggesting a critical exponent
%$x$ (if it exists indeed)
%to decrease.
For comparison, we have indicated in Fig.~\ref{fig:msdlgsim}
the predictions $x=1/2$ for single-file diffusion
and $x=2/6.25=0.32$ from the LG model.

One could
rationalize this finding as follows:
excluded-volume constraints dominate the exploration of the
small-particles' local surroundings.
In the long run, however, the
exploration of all the cul-de-sacs for the noninteracting
small particles in the frozen structure becomes vastly less effective than
an interaction-mediated transport: interacting small particles have a
larger probability to visit spaces where another small particle has just
left, thereby channeling motion along `preferential paths' known from
ion-conducting melts \cite{Horbach2002}.
%This leads to
%the crossing of MSD curves visible in Fig.~\ref{fig:msdlgsim}.

%\section{MCT Scenario}

No exact results are known for our binary mixture.
MCT describes the slow glassy dynamics well,
but a small-wave-vector divergence forbids its
application to the localization transition
\cite{Leutheusser1983,Leutheusser1983b}, and hence to a discussion of $x$.
Still, the predicted double-transition scenario
\cite{Sjoegren1986,Bosse1987,Bosse1995} is in line with our results.
%Note that a schematic model of MCT
%suppressing all wave-vector dependence yields $x=1/2$
%\cite{tvpre,Sjoegren1986}.
%
% FIG: MCT Predictions
%
\begin{figure}
\includegraphics[width=0.9\figurewidth]{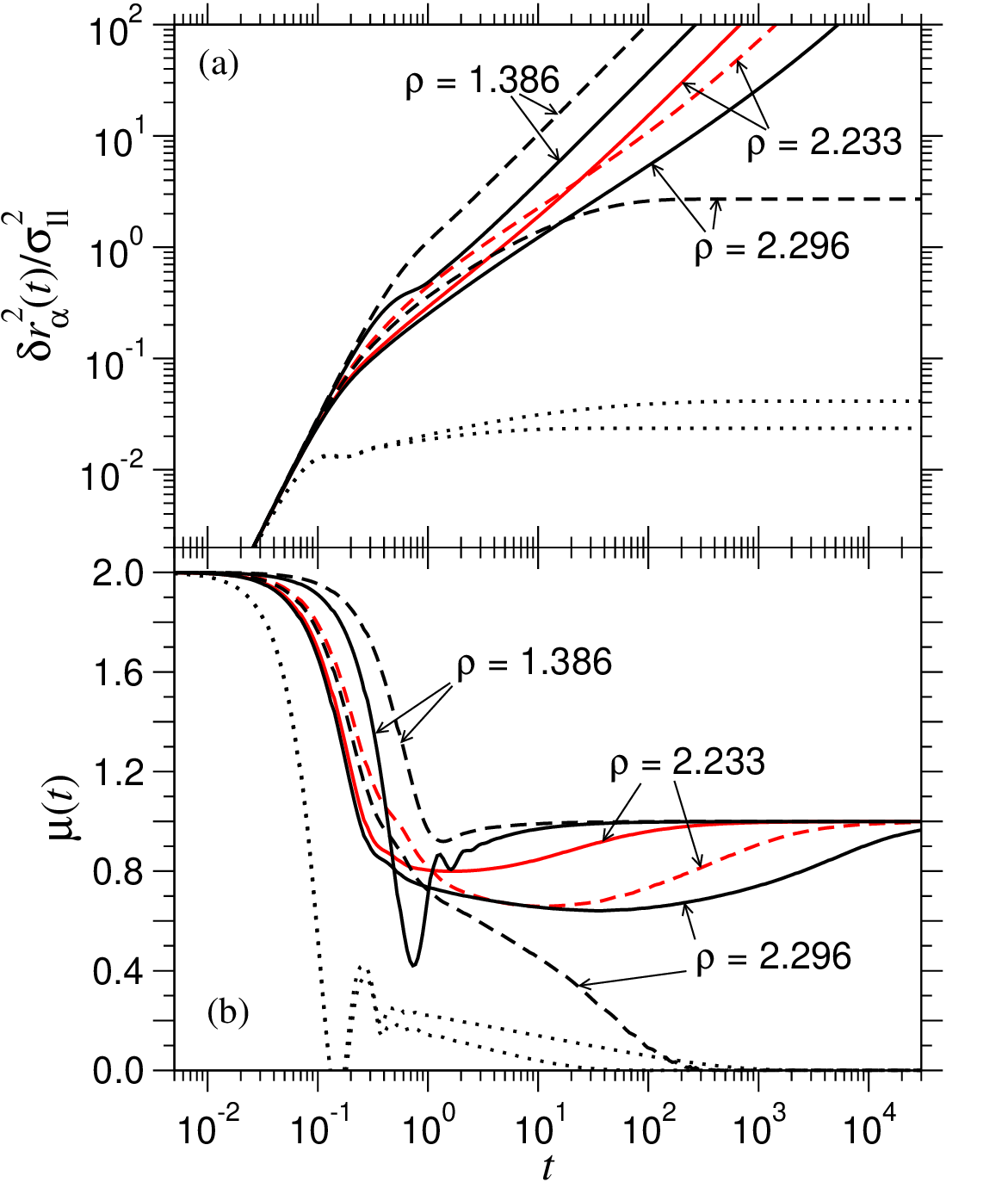}
\caption{\label{fig:mct}
  (a) Mean-squared displacements $\delta r^2_\alpha(t)$ obtained from
  mode-coupling theory with MD-simulated $S(q)$. Solid (dashed) lines:
  small-particle MSD with (without) interactions between small particles,
  densities as labeled.
  Dotted: big-particle MSD for the highest density shown.
  (b) Corresponding logarithmic derivative $\mu(t)$,
  as in Fig.~\ref{fig:msdlgsim}.
}
\end{figure}
Thus,
ignoring the asymptotic behavior close to the localization
transition, let us focus on the effect of making the small
particles `transparent' in the binary mixture.
The MCT equations for $\delta r^2_\alpha(t)$ are completely determined
once the $S_{\alpha\beta}(q,t)$ are known; we take them from MD
(cf.\ inset of Fig.~\ref{fig:diff}).
The difference in interactions enters the theory only through these quantities.
To focus on effects arising from finite wave numbers, we only show MCT
solutions
on a finite grid $q_i=(i-1/2)\Delta q\sigma_\l^{-1}$ with
$i=1,\ldots 120$, $\Delta q=0.2$ and additional low-$q$ cutoff
$q_0=4\sigma_\l^{-1}$. Discretization of $q$ in fact yields
$x_\text{MCT}=1/2$ \cite{Sjoegren1986,tvpre}.

Figure~\ref{fig:mct} shows the $\delta r^2_\alpha(t)$ for the two binary
mixtures considered. Remarkably, the theory reproduces three qualitative
trends seen in the MD data, Figs.~\ref{fig:msdsim} and \ref{fig:msdlgsim}:
(i) while the big-particle MSD exhibits the ordinary glassy two-step behavior
(localization length around $0.1\sigma_\l$),
$\delta r^2_\s(t)$ shows a different signature in the time windows accessible
to the simulation. This is the precursor of the double-transition scenario.
(ii) At low densities, transparent-small diffusion (dashed) is faster than
the one for interacting small particles (solid lines),
as expected from the reduced scattering frequency. This also holds
for high densities at intermediate times.
(iii) For large $t$ and high $\rho$,
transport of transparent small particles is much slower
than in the fully interacting case. Within the MCT picture, the latter
arises from a small shift of $\rho_c\approx2.293$ ($\gamma\approx2.9$)
to lower $\rho$, rendering transport slower at fixed $\rho$.
From our MD data, we cannot rule out whether this MCT picture is qualitatively
correct, or whether one may indeed surmise that the change in dynamics
is due to a cross-over from single-particle dynamics akin to the
Lorentz-gas model to a many-particle interaction-assisted transport.

%\section{Conclusions}

Let us summarize the main results. We studied a disparate-size
mixture of purely repulsive soft spheres
whose dynamics can be understood as the approach to two
distinct, purely dynamical arrest transitions: (i) an ordinary glass transition
connected with big-particle transport, where small-particle diffusion
does not vanish, and (ii) a localization transition
for small-particle transport at a higher density.
As a precursor, a window over increasingly large length scales appears
in the small-particle mean-squared displacement, exhibiting
power-law anomalous diffusion, $\delta r_\s^2(t)\propto t^x$.

This naturally explains an order-of-magnitude decoupling between
diffusion coefficients,
rendering our binary soft-sphere mixture a minimal model for
fast ion transport in amorphous materials.
% without resorting to Coulomb forces \cite{Maass.1991}.
Further experiments on specifically catered colloidal suspensions
\cite{Imhof1995a,Imhof1995b} would be
highly promising.

The anomalous diffusion in our binary mixture is a \emph{many-particle}
phenomenon: upon switching off interactions between the small particles,
effective power-law exponents appear to decrease.
As a consequence, excluded-volume
interactions between the small particles \emph{accelerate} their
transport in the binary mixture. This is remarkable, since in the high-density
regime one
usually expects excluded volume to hinder individual particle motion;
to understand the origin of this effect remains as a challenge for future
work.
%it is intriguing to speculate that this results from a cross-over from
%single-particle dynamics akin to the Lorentz-gas model,
%to a many-particle interaction-assisted transport.
%
%It remains a challenge to establish theoretically the critical exponents
%and universality classes of the two limiting cases.

\begin{acknowledgments}
We acknowledge discussions with T.~Franosch and F.~H\"ofling and
thoughtful comments by W.~G\"otze.
ThV is funded through the Helmholtz-Gemeinschaft, young investigators
group VH-NG 406.
\end{acknowledgments}

\bibliographystyle{apsrev}
\bibliography{lit,litnotes}

\end{document}